\documentclass[epj,referee]{svjour}

\usepackage{amssymb}
\usepackage{graphicx}

\begin{document}

\title{Empirical analysis on a keyword-based semantic system}

\author{Zi-Ke Zhang\inst{1} \and Linyuan L\"{u}\inst{1} \and Jian-Guo Liu\inst{1,2} \and Tao Zhou\inst{1,2,a}}

\institute {Department of Physics, University of Fribourg, Chemin du Mus\'ee 3, Fribourg CH-1700, Switzerland \and Department of Modern Physics,
University of Science and Technology of China, Hefei Anhui 230026, P. R. China
}

\mail{zhutou@ustc.edu}

\abstract{Keywords in scientific articles have found their significance in
information filtering and classification. In this article, we
empirically investigated statistical characteristics and
evolutionary properties of keywords in a very famous journal,
namely \emph{Proceedings of the National Academy of Science of the
United States of America} (PNAS), including frequency
distribution, temporal scaling behavior, and decay factor.
The empirical results indicate that the keyword frequency in PNAS
approximately follows a Zipf's law with exponent 0.86. In addition,
there is a power-low correlation between the cumulative number of
distinct keywords and the cumulative number of keyword occurrences.
Extensive empirical analysis on some other journals' data is also presented, with decaying trends
of most popular keywords being monitored. Interestingly, top
journals from various subjects share very similar decaying tendency,
while the journals of low impact factors exhibit completely different
behavior. Those empirical characters may shed some light on the in-depth
understanding of semantic evolutionary behaviors. In addition, the analysis of keyword-based system
is helpful for the design of corresponding recommender systems.
} 
\maketitle

\section{Introduction}
The study on semantics has a long history from its birth by Breal in
1893. It has been acquainted as a branch of
glossology. The modern semantic theory begins with the book,
\emph{Course in General Linguistics}, authorized by Saussure
\cite{Saussure1966}. As pointed out by Graemes \cite{Graemes1971},
semantics does not aim at making description of every word in the
natural language, but establishing the fundamental of descriptive
meta-language, according to which we can record and unify the
procedure of content description.

The traditional semasiology analyzes the evolutionary properties of
the acceptation mainly from the historical viewpoint, whereas the
modern theory extends the horizon to the selection of new
expressions, the existing and vanishing of phrases, systematicness
of acceptation and the meaning of sentences. Recently, as a new
interdisciplinary issue, semiotic dynamics has attracted more and more
attention from different scientific communities. Compared with the traditional
glossology and semasiology, the semiotic dynamics treats word and
morpheme as the basic unit of content, and focuses on the
understanding of how our communication pattern affects the human
semantic system, as well as the underlying mechanism of evolution,
emergence, self-organization and self-adaptation of the semantic
system \cite{Golder2006,Steels1999,Steels2006}. Therefore, semiotic
dynamics not only extends the research scope of traditional
semasiology, but also contributes to the understanding of the
characteristics of human language system, including the evolving
properties, 
competition between different
terms, the birth and fashion of new words, and so on
\cite{Nowak2002,Lieberman2007}.

The first step of the study on semiotic dynamics is to extract the
representative morphemes, such as tags and keywords of text, and
find out their relations. The mainstream methods include the Vector
Space Model (VSM) \cite{Belkin1992,Mostafa1997,Kob2000,Lin1998} and
the Ontology-Based Model (OBM) \cite{Neches1991,Steffen2003}. VSM is an
algebraic model, which describes text documents as vectors of identifiers. In VSM,
a document is represented as a vector, and each dimension corresponds to a separate term. Several methods have been
developed to calculate the values, and one of the well-known ways is the \emph{TI-IDF} weighting \cite{Salton1989}.
The weight vector for document \emph{d} can be defined as:
\begin{equation}\label{E1}
V_d=[W_1,_d,W_2,_d,...,W_N,_d]^T ,
\end{equation}
where
\begin{equation}\label{E2}
W_t,_d=tf_t\cdot\log\frac{|D|}{|{t\in{d}}|}
\end{equation}
$tf_t$ is the frequency of term $t$ in document $d$. $|D|$ is the total number of documents, and $|{t\in{d}}|$ is the number of documents which contains the term $t$. An online recommendation system \emph{Fab} \cite{Balabanovic1997} is a typical application of VSM.
However, VSM neglects the semantic content and thus its accuracy is
sensitive to the word-cutting algorithm. Comparatively,
OBM uses ontologies to describe the relationship between terms.
An ontology is a set consisted of abstracts, concepts and relations by which we wish to conceptualisze for the target world.
The most typical kind of ontology in the web has a taxonomy and an interface rules set.
The taxonomy defines the classes of terms and relations among them, while interface rules make the
terms more useful and meaningful to users \cite{Berners2001}.
An ontology-based lexical database, namely \emph{WordNet} \cite{Miller1995}, is generic ontology and free for research purposes. There are also many
limitations in the OBM, for the relations between morphemes cannot
be changed after the definition of a domain ontology
\cite{Neches1991,Steffen2003}. In addition, the keywords in the text
with special functions are usually confined within a previously
defined set of words, which update generally slower than the frontier of the
corresponding subjects.  For instance, the articles in \emph{Physical
Reviews} (A-E, L) are labeled by PACS Numbers, which can only be
selected from a standard set. The analysis on these kinds of semantic
systems can partly exhibit the correlations and statistical
evolutionary properties of keywords \cite{Liu2006,Liu2007}, however,
the establishment of this set of words involves strong external
disturbances, which hinder the understanding of essential evolving
properties driven by the semantic system itself.

There are many ways to identify the semantic characteristics of
acdemic articles which contains plentiful language signs. Thereinto the keywords, being seriously selected
by authors and/or editors, could properly represent the main content of
the corresponding article. Hence the semantic analysis of keywords
can not only avoid the above limitations in VSM and OBM, but also
shed some light on the in-depth understanding of the macroscopic
evolutionary properties of scientific activities. In this paper,
based on the data of a very famous scientific journal, we
investigated the frequency distribution,
the temporal scaling behavior, and the decay factor of keywords.

The rest of this paper is organized as follows. In section 2, we
introduce the empirical data. Section 3 shows the empirical
results, including the Zipf's plot of keyword frequency, the
temporal scaling behavior, and the decay factor. Finally, we
summarize our findings and outline some open problems of related
topics in section 4.

\begin{figure}
  \begin{center}
       \center \includegraphics[width=8cm]{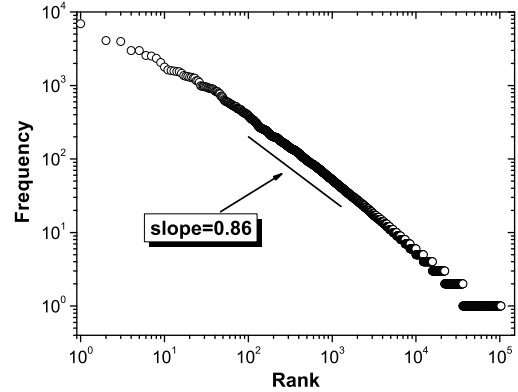}
       \caption{The frequency of keywords in PNAS follows a Zipf's law with exponent $0.86\pm0.01$.}\label{F002}
 \end{center}
\end{figure}

\section{Data}

In order to ensure the authority and representation of our
empirical analysis, we choose a journal, namely
\emph{Proceedings of the National Academy of Science of the United
States of America} (PNAS), which is a very famous scientific
journals among the world. PNAS was found in 1915 with one volume per
year. It publishes the original research articles and reports the
important academic activities. We have applied a Java script program
to automatically download the keywords of each article in PNAS from
the \emph{Web of Science}. Since the articles published from 1915 to
1990 do not have keywords, our analysis is limited in the collected
data from 1991 to 2006 (the documents without keywords, such as
\emph{Correction} and \emph{Addition}, are not considered in our
analysis), which is consisted of 46149 articles and 466470 keywords.
Those keywords are the combination of two parts: the ones added by
authors, and the ones proposed by editors (namely Keywords Plus).
Note that, some keywords are very popular and have been used in many
articles, thus the number of distinct keywords, 102992, is much
smaller than the number of keyword occurrences (i.e., 466470).
Hereinafter, when referring to the number of keywords, we mean the
total number of keyword occurrences. For example, if there are two
articles, one has keywords A, B and C, while the other has keywords
C and D. Then, we say there are 5 keywords, and 4 distinct keywords.
Data of some other journals are also analyzed for comparison (see
below). To be comparable, we also extract the data from 1991 to
2006.

\begin{figure}
  \begin{center}
       \center \includegraphics[width=8cm]{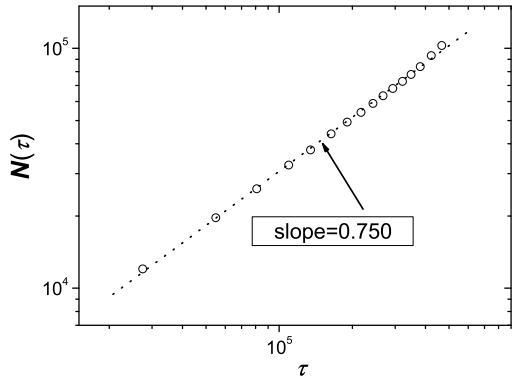}
       \caption{The scaling relation between $\tau$ and $N(\tau)$. The 16 points, from left to right, represent the cumulative data. That is to say, the leftmost point corresponds to the cumulative value up to the year 1991, the second point from left denotes the cumulative value up to the year 1992, etc. }\label{F002}
 \end{center}
\end{figure}

\section{Statistical Analysis}
\subsection{Zipf's law of keywords' occurrences}
In 1930s, Zipf put forth a famous law of frequency distribution of
phrases, namely \emph{Zipf's Law} \cite{Zipf1932}, which has
been widely used to characterize the distributions of firm size
\cite{Stanley1995,Axtell2001}, city scale \cite{Rosen1980}, wealth
\cite{Levy1997,Xie2005}, earthquake strength \cite{Bak1989}, and so
on. Rank the phrases in a descending order according to their
occurring frequency in a text, Zipf found a power-law relation
between the rank, $n$, and its corresponding frequency, $P_n$, as:
\begin{equation}
P_{n}=n^{-\alpha}.
\end{equation}
As shown in Fig. 1, the frequency distribution of keywords in PNAS
approximately follows a Zipf's law with exponent 0.86 crossing 4
magnitudes. Most keywords are of low frequencies, while a few popular keywords appear very frequently. Up to
2006, the most popular keyword, \emph{Expression}, has been used for
6927 times. Meanwhile, there are 66782 (64.84\%) distinct keywords used only
once. As shown in Table 1, this Zipf's law is universally existed
for various scientific journals in different subjects.

\subsection{Scaling between the number of distinct keywords and the total number of keywords}
A keyword in a new publication is either a new one or has appeared
in a prior article. Denote $\tau$ the cumulative number of keywords,
and $N(\tau)$ the corresponding cumulative number of distinct
keywords. Figure 2 presents a power-law relation between $\tau$ and
$N(\tau)$ during the evolving process from the year 1991 to 2006.
The dash line, with slope $0.750\pm0.007$, is the linear fitting of
the data, that is to say,
\begin{figure}
  \begin{center}
       \center \includegraphics[width=9cm]{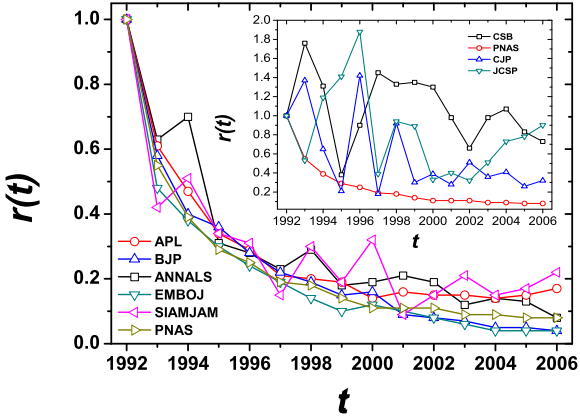}
       \caption{(Color online) The decay factor $r_{t}$ as a function of time (year resolution) for different journals. The inset compares PNAS and several local journals with much lower impact factors. The full titles of the journals can be found in Table 1.}\label{F002}
 \end{center}
\end{figure}

\begin{equation}\label{E3}
N(\tau)=c\tau^{\lambda},
\end{equation}
where $\lambda\approx0.75$, and $c$ is a constant. From Eq.
(\ref{E3}), one can derive that the growing rate of distinct keywords
is $c\lambda\tau^{\lambda-1}$, where $\tau$ is the number of
keywords. When $\lambda=1$, there exists a linear relation between
the number of newly added distinct keywords and that of the newly
added keywords, and the growing rate is a constant $c$. When
$\lambda<1$, the growing rate of distinct keywords will decrease
with the increasing of the total number of keywords. Actually, if
the number of distinct keywords is $N$, the probability that the
next keyword has not been used before (i.e., distinct) is equal to
$c^{2\lambda-1}\lambda N^{1-\frac{1}{\lambda}}$, which will
decreases with the increase of $N$ when $\lambda<1$. The data from
some other scientific journals indicate the universality of this
scaling law (see Table~1).

Surprisingly, some recently empirical studies demonstrate the
extensive existence of this kind of scaling law, with the same form as Eq.
(4), in the web tag systems
\cite{Cattuto20071,Cattuto20072,Halpin2007}. Note that, the
collaborative tagging system is an open and optional system where
each user can optionally modify the tags in the system. In contrast,
the keywords in articles are considered seriously by the
authors and editors, so the keyword-based systems are more canonical
and serious. However, both tags and keywords follow the same scaling
law. This result indicates a possibly universal law for the generic
semantic systems.

\begin{table}
 \caption{Statistics of several journals from different
subjects, including Appl. Phys. Lett. (APL), British J. Pharmacology
(BJP), EMBO J. (EMBOJ), Annals of Neurology (AN), SIAM J. Appl.
Math. (SIAM), Chin. Sci. Bull (CSB), CZECH. J. PHYS. (CJP), J. CHEM.
SOC. PAKISTAN (JCSP). $\alpha$ is the exponent in the Zipf's plot
and $\lambda$ is the scaling exponent defined in Eq. (4). IF stands for the
impact factor of the journal in 2007.}
\begin{center}  \begin{tabular} {cccc}
  \hline \hline
   Journal Title     & IF        &  $\alpha$  &     $\lambda$  \\ \hline
   APL & 3.596     & 1.01$\pm$0.01  & 0.683$\pm$0.008     \\ 
   BJP & 3.767 &  0.92$\pm$0.01  & 0.753$\pm$0.006   \\ 
   PNAS & 9.598                         & 0.86$\pm$0.01  & 0.750$\pm$0.007    \\ 
   EMBOJ & 8.662              & 0.86$\pm$0.01  & 0.753$\pm$0.003    \\ 
   AN & 8.813  & 0.83$\pm$0.02  & 0.716$\pm$0.005    \\ 
   SIAM & 1.026 & 0.58$\pm$0.02  & 0.825$\pm$0.005    \\ 
   CSB & 0.77        & 0.51$\pm$0.01  & 0.857$\pm$0.013     \\ 
   CJP & 0.423       & 0.48$\pm$0.01 & 0.912$\pm$0.002    \\ 
   JCSP &0.095       & 0.39$\pm$0.01 & 0.916$\pm$0.004    \\ 
   \hline \hline
    \end{tabular}
\end{center}
\end{table}

\begin{figure}
  \begin{center}
       \center \includegraphics[width=8cm]{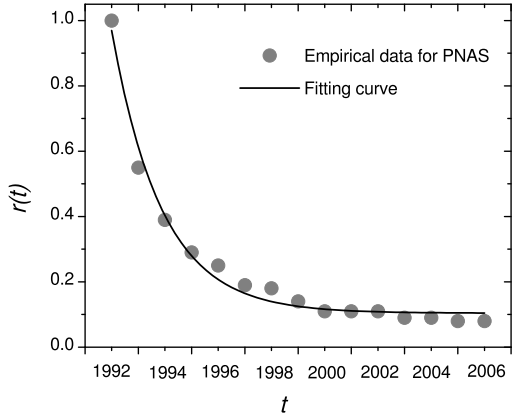}
       \caption{The decaying trend of top-10 keywords in the year 1991 for PNAS. The circles represent empirical result obtained by Eq. (6), while the solid curve corresponds to the fitting function shown as Eq. (7).}
 \end{center}
\end{figure}

\subsection{Decaying behavior of the most popular keywords}
The decay factor $r_{t}$ of a keyword describes the collective decay
of attention, which can be defined as \cite{Wu2007}:
\begin{equation}
r_{t}=\frac{\texttt{log}N_t-\texttt{log}N_{t-1}}{\texttt{log}N_1-\texttt{log}N_0},
\end{equation}
where $N_{t}$ denotes the cumulative occurring frequency of the monitored
keyword at time $t$ with year resolution. $N_{0}$ is the occurring
frequency in the first year (i.e., the year 1991). In order to
reduce the fluctuation, when analyzing the decay factor, we
use the aggregated data of several keywords, thus the Eq. (5) should
be rewritten as:
\begin{equation}
    r_{t}=\frac{E(\texttt{log}N_{t})-E(\texttt{log}N_{t-1})}{E(\texttt{log}N_{1})-E(\texttt{log}N_{0})},
\end{equation}
where $E(\cdot)$ denotes the average over the monitored set of
keywords. We analyze the decay factor of ten most popular keywords
(top-10 keywords for short) in the year 1991. As shown in Fig. 3,
$r_{t}$ of PNAS (red circles) decays very fast in the first three
years, and then slows down. The decay factor almost decreases to a
half in 1993. Actually, its decaying trend can be well fitted by
an exponential function as
\begin{equation}
 y= A_1 * e^{-x/t_1}+ y_0,
\end{equation}
where $y_0$= 0.10 $\pm$ 0.01, $A_1$= 1.47 $\pm$ 0.09, $t_1$= 1.88
$\pm$ 0.13, and the time $x$ varies from 1 (the year 1992) to 15
(the year 2006). The fitting curve versus empirical result is shown
in Fig. 4. This decaying trend can be used to quantify the broadness
of interests of a journal. For a journal with high impact factor, it
is possible and reasonable that $r_t$ decays very fast in the early
stage since it mainly publishes the newest progress in natural
science with some new concepts.

We also empirically study the decaying behavior of top-10 keywords for
several top journals in different subjects, from biology to
mathematics. As shown in Fig. 3, all those decaying curves display
similar tendency. In contrast, as shown in the inset of Fig. 3,
$r_t$ of three local journals with relatively lower
scientific impacts have far different shapes compared with those top
journals. Actually, the decay factor $r_t$ exhibits large fluctuation and no obvious decaying tendency can be observed even in a long period of time (15 years). A possible reason is that those journals with low impacts do not publish as many newest progresses as top journals.

\section{Conclusion and Discussion}
In this paper, we empirically investigated the statistical
characteristics and the evolutionary properties of keywords in a
very famous journal, namely PNAS, including the frequency
distribution, the temporal scaling behavior, and the decay factor.
Firstly, the empirical results indicate that the frequency distribution of
keywords in PNAS approximately follows a Zipf's law with exponent
0.86, which means only a few keywords are used frequently in PNAS,
whereas most of keywords are used unusually. Secondly, there is a
power-low correlation between the number of distinct keywords and
the total number of keyword occurrences. We have also investigated
the data of some other journals in different subjects, which
strongly indicate the universality of those two statistical
properties. In addition, we studied the decay factor of the most
popular keywords. Interestingly, the top journals, though from far
different subjects, exhibit very similar decaying behavior that can be approximately fitted by an exponentional function. While
the journals with lower impact factors exhibit very different behaviors,
actually, no obvious decaying tendency is observed.

The studies of systems with collaborative keywords are also relevant
to the recent progress on the design of recommender systems.
Actually, with the advance of \emph{Web2.0} technique, a great
number of recommendation algorithms were applied to some on-line
resource-sharing systems \cite{Adomavicius2005}, which can recommend
music, films, books and news to users according to their historical
activities. Up to now, the most accurate algorithm is content-based
\cite{Pazzani2007}. However, those content-based methods are
practical only if the items have well-defined attributes, which can be extracted automatically. The traditional content
analyzing approach, based on cutting the content word by word, is
often impractical since its computational complexity is too high for
the huge-size database. In contrast, the structure-based algorithm
has less complexity but also lower accuracy
\cite{Zhou2007,Zhang2007,Zhou2008}. Because the keywords of an article
can express, to some extent, the main content of this article, an
algorithm with low complexity and high accuracy is expected by
properly integrating the recommendations drawn from the keyword-article bipartite graph
and the author-article bipartite graph (see Ref. \cite{Zhou2007} how to get recommendations from a bipartite graph).

In addition, a Keyword-Based Collaboration Network (KBCN) can be
constructed based on the definition that two keywords are connected
if they appeared together in at least one article. More characters
about the structural organization of a keyword-based semantic system
can be analyzed with the help of KBCN (see Refs.
\cite{collaboration1,collaboration2,collaboration3} how to construct
and analyze collaboration networks). Especially, the in-depth
understanding of the hierarchical organization \cite{Ravasz2003},
the community structure \cite{Girvan2002} and the motif density
\cite{Xiao2007} are crucial for the classification of research areas
and the evaluation on the strength of interdisciplinary studies.

\section{Acknowledgement}
This work is supported by SBF (Switzerland) for financial support
through project C05.0148 (Physics of Risk), and the Swiss National
Science Foundation (Project 205120-113842). T.Z. acknowledges the
National Natural Science Foundation of China (Grant Nos. 10635040
and 60744003).

\end{document}